\documentclass{PoS}
\usepackage{setspace}
\usepackage{graphicx}
\usepackage{dcolumn}
\usepackage{multirow}
\usepackage{subfigure}
\usepackage{times,mathptm}
\usepackage{float}
\usepackage{color}
\usepackage{cite}
\usepackage{amsmath,amsfonts}
\usepackage{mathptmx}
\usepackage{mathrsfs}
\usepackage{bbm}
\usepackage{bm}
\usepackage{xfrac}

\newcommand{\beq}{\begin{equation}}
\newcommand{\eeq}{\end{equation}} 
\newcommand{\bea}{\begin{eqnarray}}
\newcommand{\eea}{\end{eqnarray}}

\newcommand{\tK}{\widetilde{K}}
\newcommand{\p}{\phi}
\renewcommand{\b}{\beta}
\renewcommand{\a}{\alpha}

\newcommand{\tr}{\text{Tr}}

\newcommand{\bx}{\mathbf{x}}
\newcommand{\by}{\mathbf{y}}
\newcommand{\bk}{\mathbf{k}}

\renewcommand{\p}{\psi}
\newcommand{\pb}{\overline{\psi}}
\newcommand{\vx}{\bx}
\newcommand{\vy}{\by}

\newcommand{\vk}{\bk}

\newcommand{\m}{\mu}

\newcommand{\D}{\Delta}

\renewcommand{\th}{\theta}
\renewcommand{\p}{\psi}

\newcommand{\oh}{\frac{1}{2}}

\newcommand{\dg}{\dagger}
\newcommand{\non}{\nonumber}

\newcommand{\pa}{\partial}

\title{Relative weights approach to dynamical fermions at finite densities}

\ShortTitle{Relative weights approach to dynamical fermions at finite densities}

\author{\speaker{Jeff Greensite}
        \thanks{Work supported by the US Department of Energy under Grant No.\ DE-SC0013682.}\\
       San Francisco State University\\
       E-mail: \email{greensit@sfsu.edu}}

\author{Roman H\"ollwieser
           \thanks{Work supported by the
        Austrian Research Fund (FWF) under Contract No. J3425-N27.}\\
        New Mexico State University\\
        E-mail: \email{roman.hoellwieser@gmail.com}
}
\abstract{The method of relative weights, coupled with mean field theory, is applied to the problem of simulating
gauge theories with dynamical staggered fermions at finite densities.  We present initial results and discuss issues so far encountered.}

\FullConference{34th annual International Symposium on Lattice Field Theory\\
                 24-30 July 2016\\
                 University of Southampton, UK}

\begin{document}

    The relative weights approach to the sign problem in gauge field theories \cite{Greensite:2013yd,Greensite:2013bya} is an approach which maps the original lattice theory into an effective Polyakov line action (PLA), and then solves the effective theory at finite chemical potential via mean field theory \cite{Greensite:2014isa,Greensite:2014cxa}.  The PLA 
$S_P$ is defined as the theory which remains after integrating out all gauge and matter degrees of freedom, subject to the constraint that the Polyakov line holonomies are held fixed.  This is most convenient in temporal gauge
\bea
\exp[S_P[U_{\bx}]] &=&  \int  DU_0(\bx,0) DU_k  D\pb D\p   \left\{\prod_\bx \delta [U_{\bx}-U_0(\bx,0)]  \right\} e^{S_L} \ .
\eea
Given $S_P$ at $\m=0$, the action at finite $\m$ is simply
\bea
     S_P^\m[U_\vx,U^\dg_\vx] =  S_P^{\m=0}[e^{N_t \m} U_\vx,e^{-N_t \m}U^\dg_\vx]   \ .
\eea
For heavy quarks, the PLA can be derived via the hopping parameter expansion \cite{Fromm:2011qi}. We are interested in going to lighter quark masses, where that method cannot be easily applied.

    Although it is difficult to compute $S_P$ directly, it is straightforward to compute derivatives of $S_P$ with respect to small variations
of the Polyakov line holonomies.  Let $S'_L, S''_L$ be the lattice actions in temporal gauge with $U_0(\vx,0)$ fixed to $U'_\vx, U''_\vx$
respectively, and $\D S_P = S_P(U'_\vx) - S_P(U''_\vx)$.  Then
\bea
\exp[\D S_P] &=& {\int  DU_k  D\pb D\p~  e^{S'_L} \over \int  DU_k  D\pb D\p ~  e^{S''_L} }
\non \\
&=&{\int  DU_k  D\pb D\p ~  \exp[S'_L-S''_L] e^{S''_L} \over \int  DU_k  D\pb D\p ~  e^{S''_L} }
\non \\
&=& \Bigl\langle  \exp[S'_L-S''_L] \Bigr\rangle''  \ ,
\eea
where $\langle ... \rangle''$ means the VEV in the Boltzman weight $\propto e^{S''_L}$.  Now suppose $U_\vx(\lambda)$ is some path through configuration space parametrized by $\lambda$, and suppose $U'_\vx$ and $U''_\vx$ differ by a small change in that parameter, i.e.  $U'_\bx = U_\bx(\lambda_0 + \oh \D \lambda) ~,~ U''_\bx = U_\bx(\lambda_0 - \oh \D \lambda )$.
Then the relative weights method gives us the derivative of the true effective action $S_P$ at any point along the path:
\beq 
\left( {dS_P \over d\lambda} \right)_{\lambda=\lambda_0}  \approx  {\D S_P \over \D \lambda} \ .
\eeq
We find it useful to take derivatives with respect to Fourier components $a_\vk$ of Polyakov lines $P_\vx$, where
$P_\bx \equiv {1\over N_c} \tr U_\vx = \sum_\bk a_\bk e^{i\bk \cdot \bx}$. 
We first set a particular momentum mode $a_\vk$ to zero.  Call the resulting configuration 
 $\widetilde{P}_x$. Then define ($f \lesssim 1$)
\bea
            P''_\vx &=& \Bigl(\a - \oh \D \a \Bigr) e^{i\vk \cdot \vx} + f \widetilde{P}_x
\non \\
            P'_\vx &=& \Bigl(\a + \oh \D \a \Bigr) e^{i\vk \cdot \vx} + f \widetilde{P}_x \ ,
\eea           
which uniquely determine (in SU(2) and SU(3)) the eigenvalues of the corresponding holonomies 
$U'_\bx,U''_\bx$.  In this way we can compute
\beq
{1\over L^3}\left( {\pa S_P \over \pa a_{\vk}}\right)_{a_\vk = \a}  \ .
\eeq

\begin{figure}[h!]
\centering
\subfigure[]  
{   
 \label{ks504}
 \includegraphics[scale=0.5]{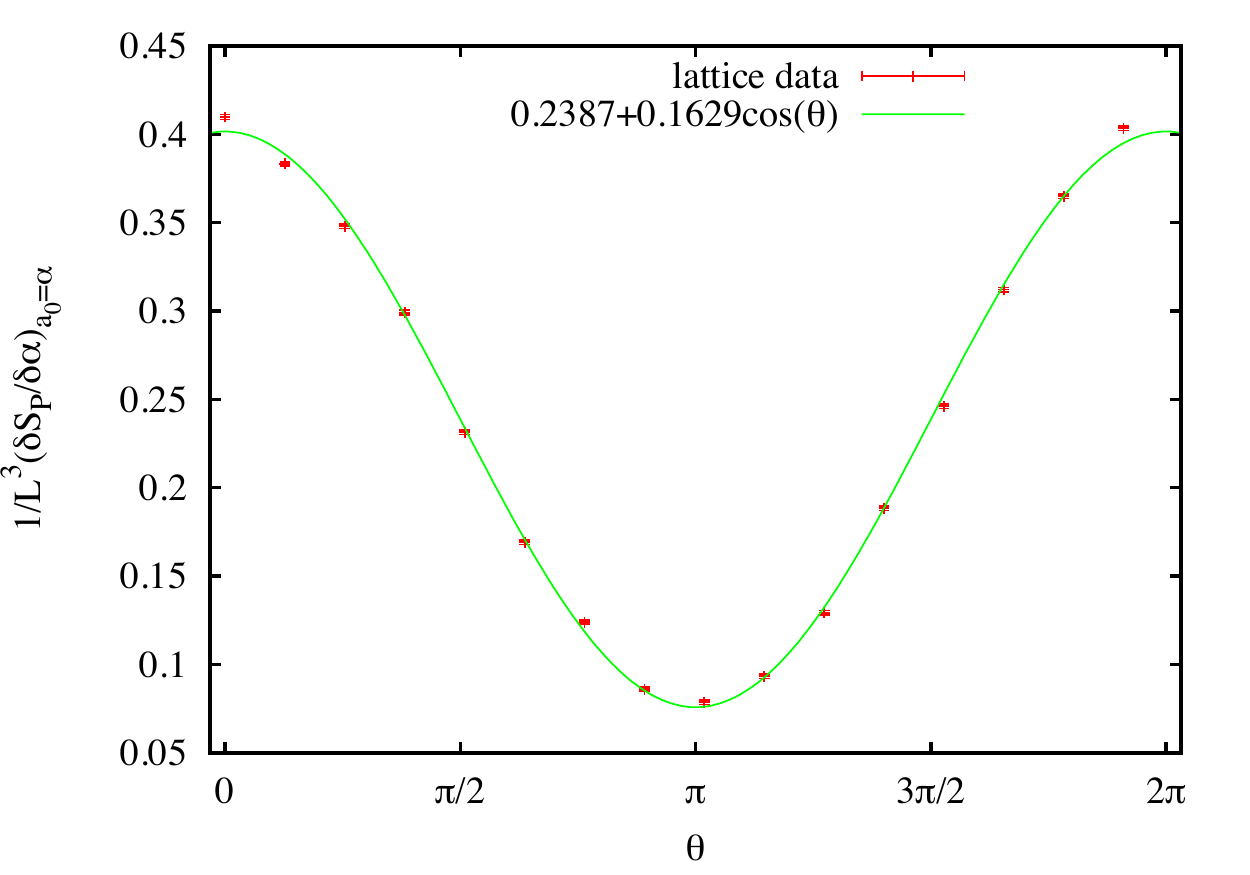}
}
\subfigure[]  
{   
 \label{kk504}
 \includegraphics[scale=0.5]{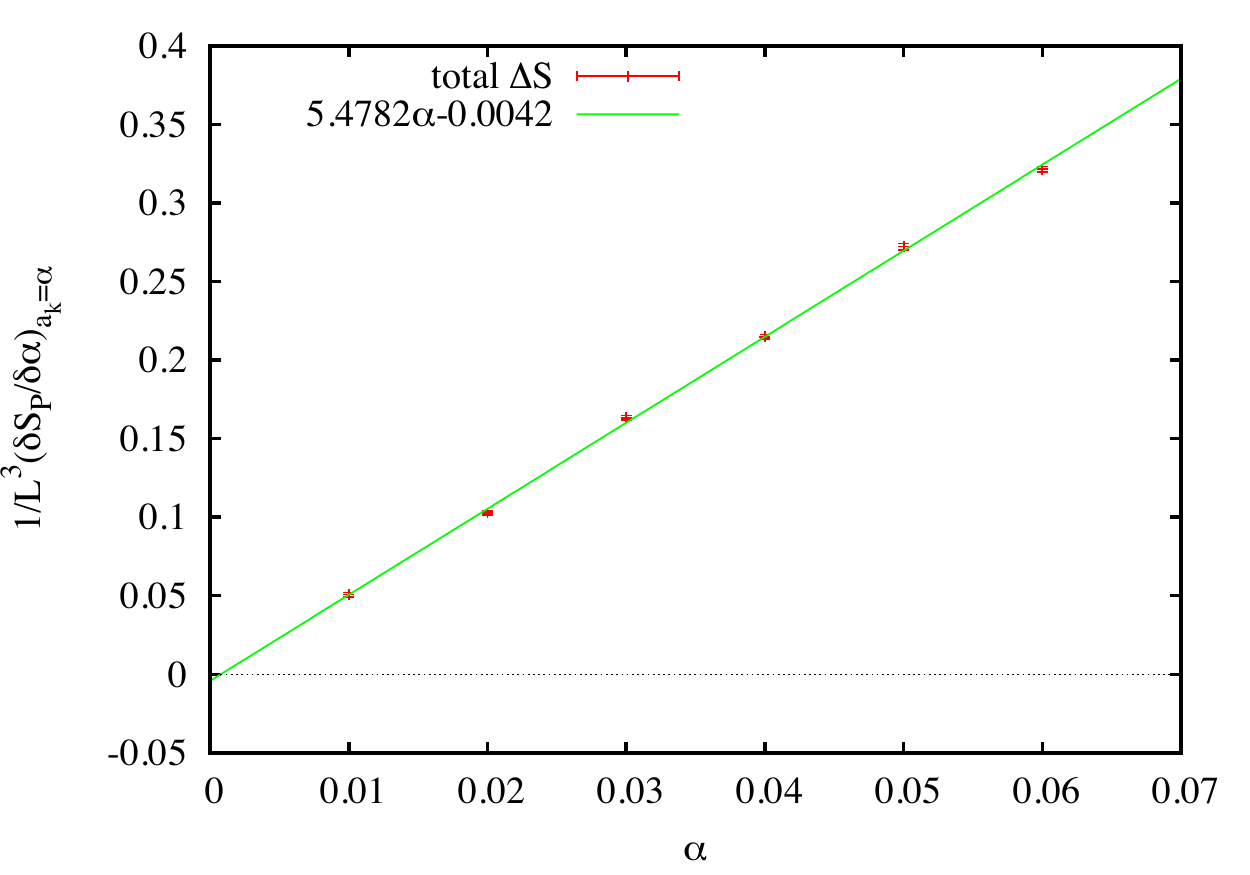}
}
\caption{SU(3) Wilson action, staggered fermions, $\beta=5.2,~ma=0.35,~N_t=4$.  (a) $S_P$ derivative with respect to the zero mode, vs.\ imaginary chemical potential $\m/T = i\theta$ at $a_0=0.03.$ (b)  $S_P$ derivative wrt $a_k$ with $k_L \ne 0$ (mode numbers (2,1,0)), evaluated at various $a_k=\a$.} 
\label{dS}
\end{figure}

    Motivated by the known contribution of heavy-dense quarks to the effective action \cite{Fromm:2011qi} (see also \cite{DePietri:2007ak})  we fit the relative weights data for the derivatives of $S_P$ to the ansatz 
\bea
e^{S_P} &=& \prod_\vx \det[1+he^{\mu/T}\tr U_\vx] \det[1+he^{-\mu/T}\tr U^\dg_\vx] 
  \exp\left[\sum_{\vx,\vy} P_\vx K(x-y) P^\dg_\vy \right] 
\eea   
for staggered, unrooted quarks to determine the parameter $h$ and kernel $K(\vx-\vy)$. Of course this ansatz is not exact.  An important check is to compute and compare, at $\m=0$, the Polyakov line correlator $G(|\vx-\vy|) = \langle P(\vx) P^\dg(\vy) \rangle$ 
in both the PLA, and the underlying lattice gauge theory.   We gain precision by introducing an imaginary chemical potential $\m/T = i\theta$.  Construct $U'_\vx, U''_\vx$ as before, then set $U'(\vx,0) = e^{i\theta} U'_\vx ~,~ U''(\vx,0) = e^{i\theta} U''_\vx$.  To lowest order in $h$, we then have
\bea
{1\over L^3} \left( {\pa S_P \over \pa a_0} \right)_{a_0=\a}^{\m/T=i\theta}  = 2 \tK(0) \a + 6 h \cos \theta \ ,
\eea
where $\tK(\vk)$ is the Fourier transform of $K(\vx)$.  Fitting the data for the left hand side at various $\th$  determines $h$ and $\tK(0)$.
Likewise, at $\vk \ne 0$ at lowest order in $h$
\bea
 {1\over L^3} \left( {\pa S_P \over \pa a^R_\vk} \right)_{a_\vk=\a} = 2 \tK(\vk) \a  \ .
\eea
From this we can determine $\tK(\vk)$.  Sample data and fits are shown in Fig.\ \ref{dS}.
 
   As in previous work with bosonic matter fields, we fit the data for $\tK(\vk)$ by two straight lines
\bea
            \tK^{fit}(\bk) = \left\{ \begin{array}{cl}
                   c_1 - c_2 k_L & k_L \le k_0 \cr
                   d_1 - d_2 k_L & k_L \ge k_0 \end{array} \right. \ ,
\eea
where
\bea
           k_L = 2 \sqrt{ \sum_{i=1}^3 \sin^2(k_i/2) } 
\eea
is the lattice momentum.  The last few points at lowest momenta, which do not fit on a straight line, are handled by a long distance cutoff.  Define
\bea
    K(\vx-\vy) = \left\{ \begin{array}{cl}
                   {1\over L^3}\sum_\vk \tK^{fit}(k_L) e^{i\vk\cdot (\vx-\vy)} & |\vx-\vy| \le r_{max} \cr \\
                      0 & |\vx-\vy| > r_{max} \end{array} \right. \ ,
\eea
and Fourier transform again to $\tK(k)$.  The parameter $r_{max}$ is chosen to give a best fit to the data points at lowest
momenta. 
 
\begin{figure}[h!]
\centering
\subfigure[~$\beta=5.04,ma=0.2$]  
{   
 \label{pl504}
 \includegraphics[scale=0.4]{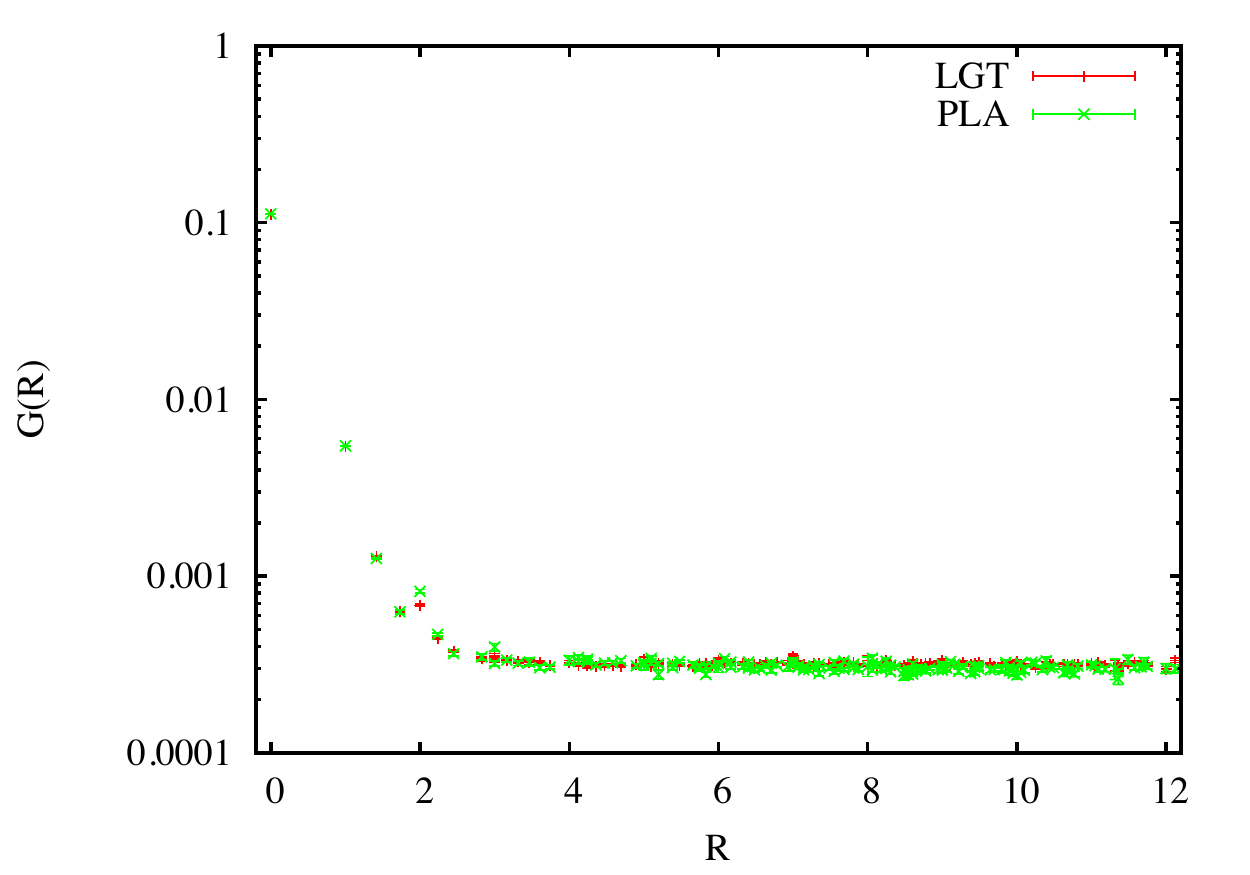}
}
\subfigure[$\beta=5.2, ma=0.35$]  
{   
 \label{pl52}
 \includegraphics[scale=0.4]{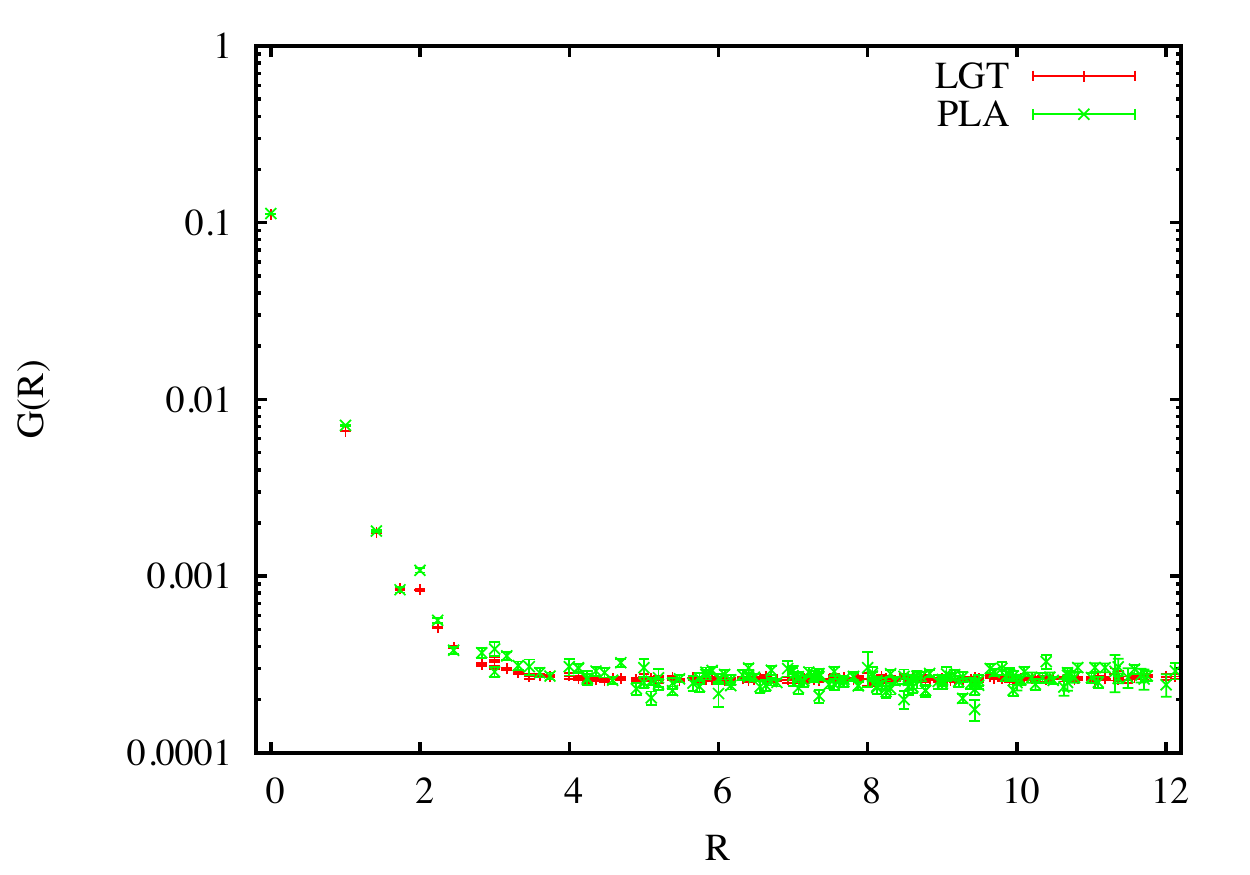}
}
\subfigure[$\beta=5.4,ma=0.6$]  
{ 
 \label{pl54}
 \includegraphics[scale=0.4]{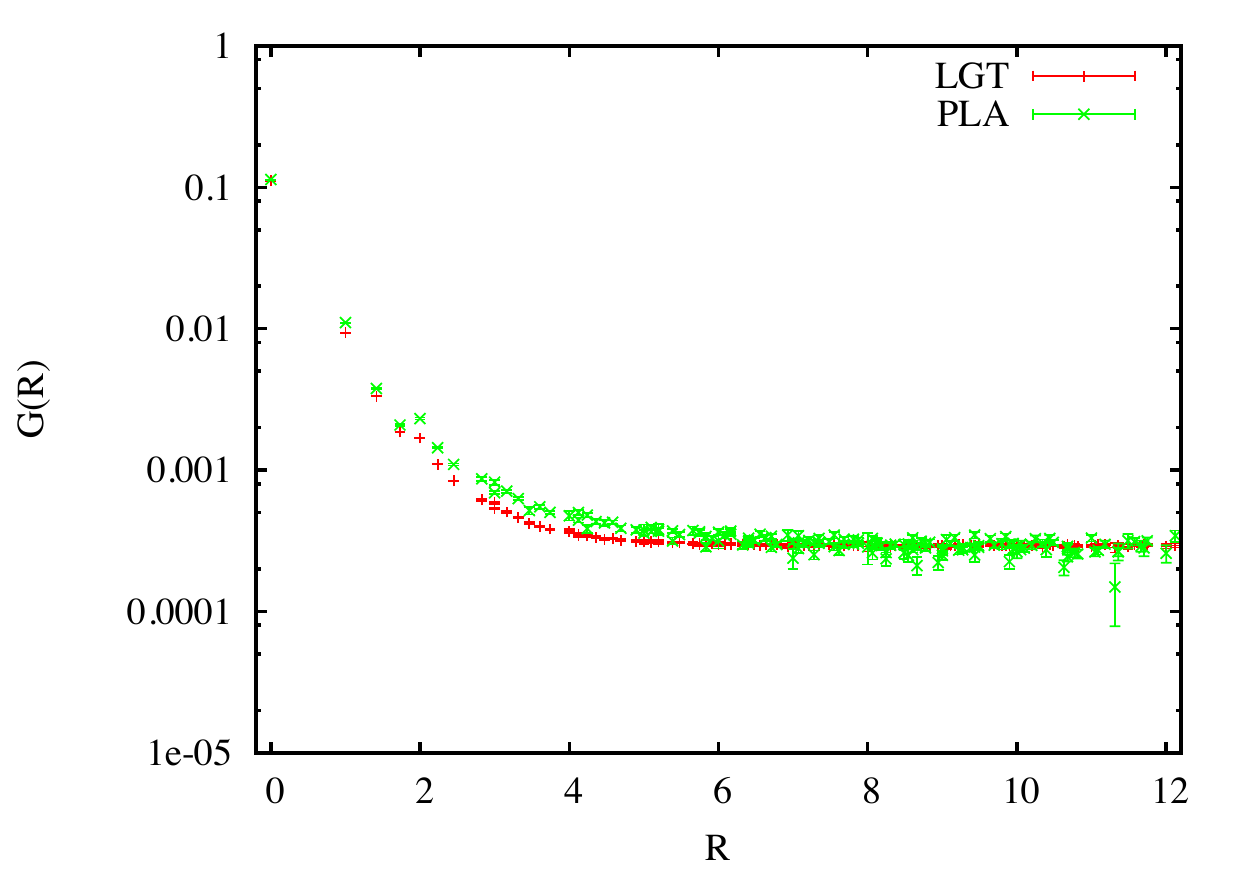}
}
\caption{Comparision of Polakov line correlators in the PLA and in the underlying gauge theory (Wilson action) at $N_t=4$, for three choices of coupling and quark mass.}
\label{plcorr}
\end{figure}

    Using this method to determine the parameter $h$ and kernel $K(|\vx-\vy|$, the effective theory is defined.  We can then compute the Polyakov line correlator $G(R)$ at
$\m=0$ in both the PLA and also in the underlying lattice gauge theory, and check to see to see if they agree.  The results for
SU(3) gauge theories with dynamical staggered quarks, no rooting, with the Wilson action on a $16^3 \times 4$ lattice volume and various gauge couplings and quark masses are shown in Fig.\ \ref{plcorr}.  In these cases the agreement seems excellent.  Note that the quark masses are far out of the range of validity of a strong-coupling/hopping parameter expansion.

\begin{figure}[h!]
\centering
 \includegraphics[scale=0.5]{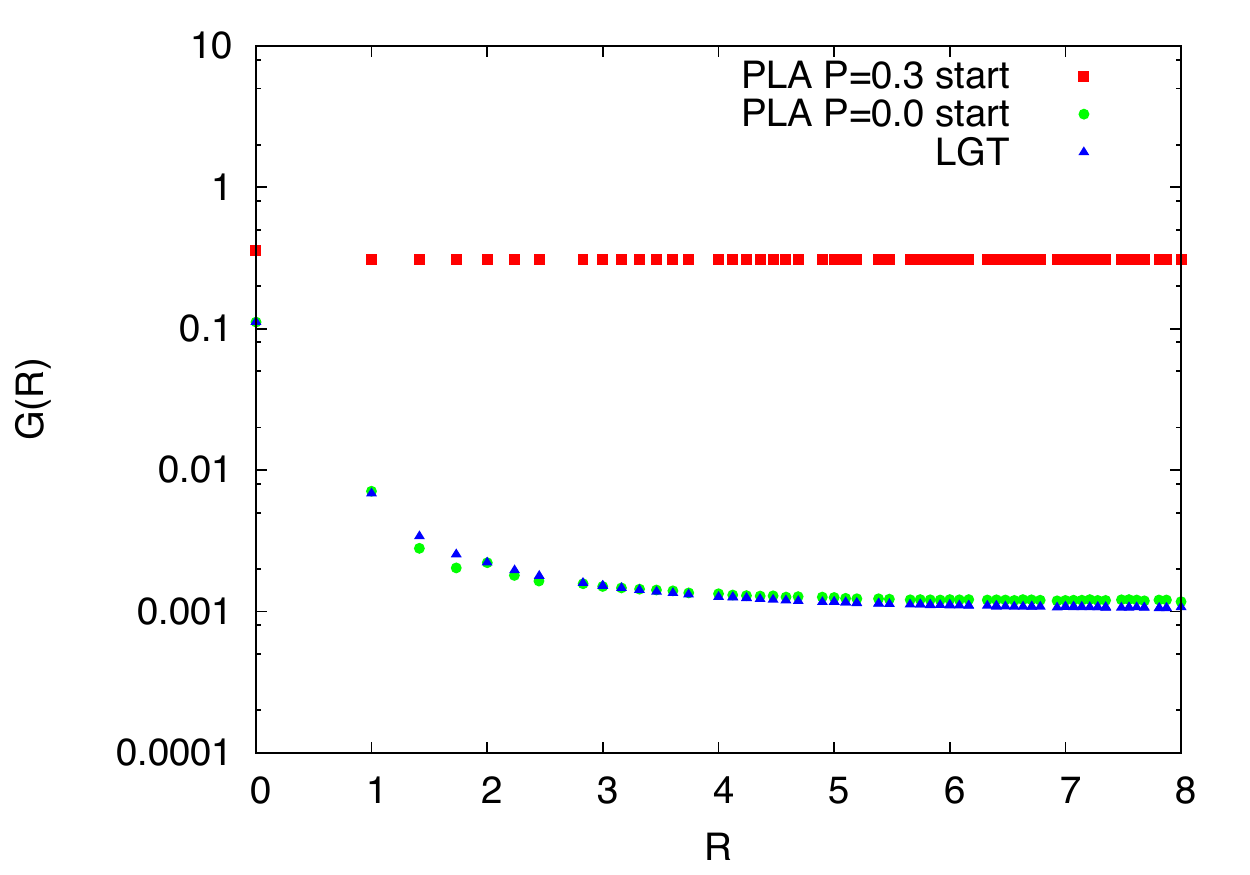}
\caption{Comparision of Polakov line correlators in the PLA and in underlying gauge theory (L\"uscher-Weisz gauge action, $\beta=7.0,~ma=0.3,~N_t=6$).  Note that the answer for the PLA depends on the initialization, which implies the existence of very long-lived metastable states.}
\label{LW}
\end{figure}

    We also tried the L\"uscher-Weisz SU(3) gauge action at $\beta=7.0, ma=0.3, N_t=6$.  Unlike previous cases, the couplings in
the effective action are completely non-local: all spins are coupled to all other spins, at least on a $16^3$ lattice.
In this instance we found that the simulation of the PLA depends on the starting point; i.e.\ there are long-lived
metastable states, persisting for many thousands of sweeps (see Fig.\ \ref{LW}).  A start with $P_\vx=0$ seems to choose the phase which agrees with underlying lattice gauge theory, while e.g.\ initialization at $P_\vx=0.3$ does not.   This introduces an unfortunate ambiguity:  how do we pick the correct phase at $\m \ne 0$?  It should be emphasized that this is not a question of having a sign problem.  The problem in this case arises even at $\m=0$, and is presumably connected with the highly non-local all-spins-coupled-to-all-spins character of the effective action.  It may be necessary to restrict the investigation to a parameter range where this issue does not arise.

    Leaving this ambiguity aside for the moment, and having arrived at an effective PLA, there is still a sign problem.  We deal with this via mean field theory, as discussed in \cite{Greensite:2012xv}, and in this case the presence of many spin couplings beyond nearest neighbor is actually a great advantage.  Mean field methods were applied to such models at $\m \ne 0$ in \cite{Greensite:2014cxa}, and the results were compared to results obtained in the Langevin approach. It was found in that case that the mean field and Langevin methods agree perfectly, except where the Langevin method fails due to the singular drift problem pointed out by Mollgaard and Splittorff \cite{Mollgaard:2013qra}.  So this is the method we apply to solve the PLA at $\m=0$; details of the method can be found in \cite{Greensite:2014cxa}.

\begin{figure}[h!]
\centering
\subfigure[]  
{   
 \label{uv504}
 \includegraphics[scale=0.4]{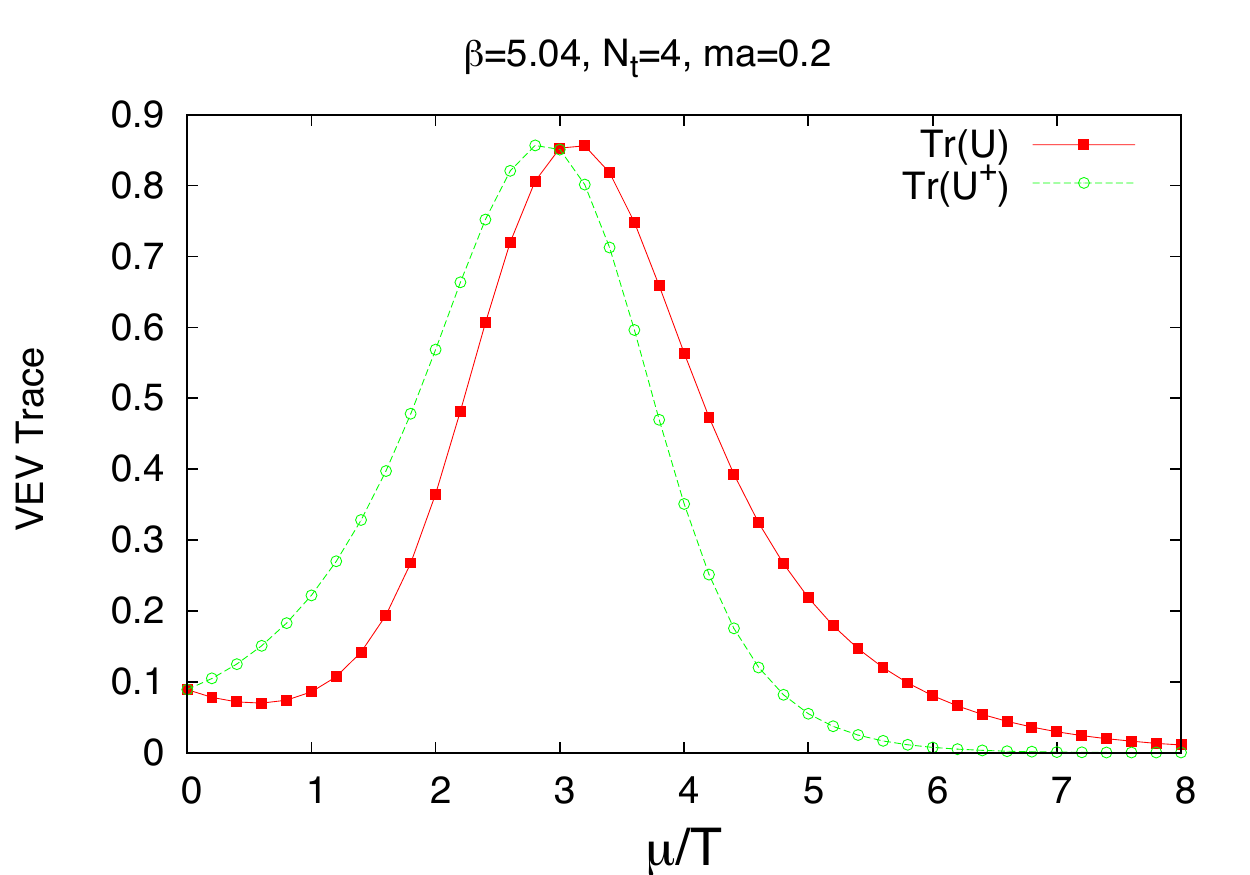}
}
\subfigure[]  
{   
 \label{n504}
 \includegraphics[scale=0.4]{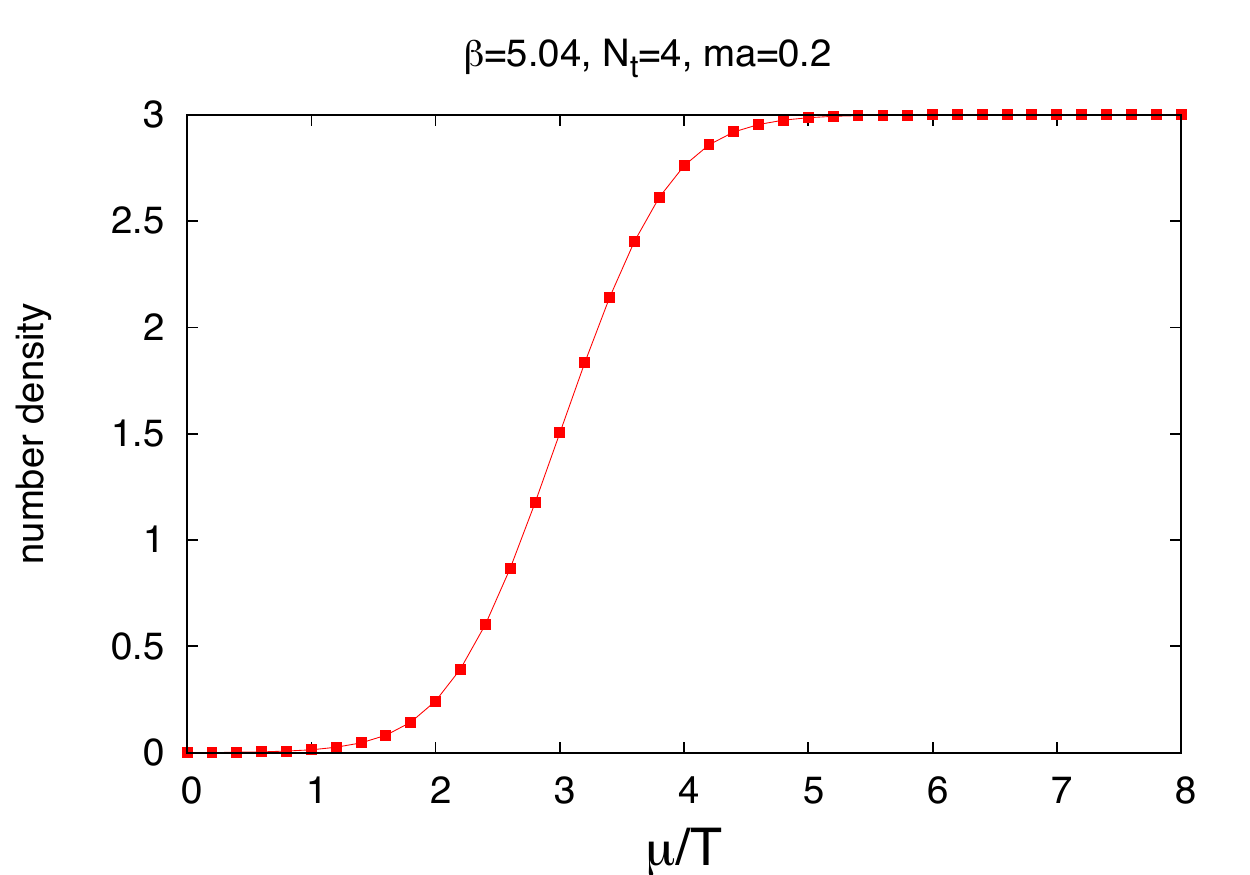}
}
\subfigure[]  
{   
 \label{uv54}
 \includegraphics[scale=0.4]{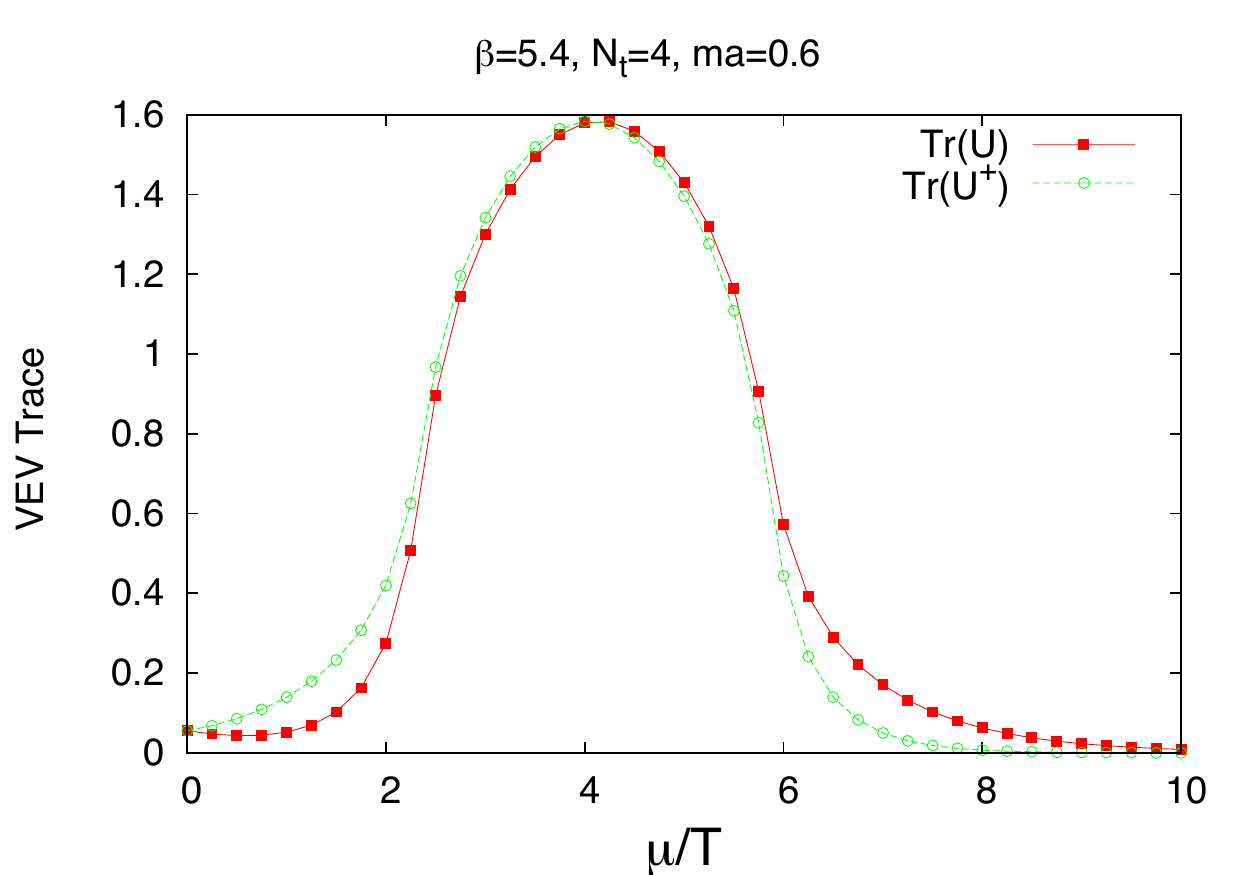}
}
\subfigure[]  
{   
 \label{n54}
 \includegraphics[scale=0.4]{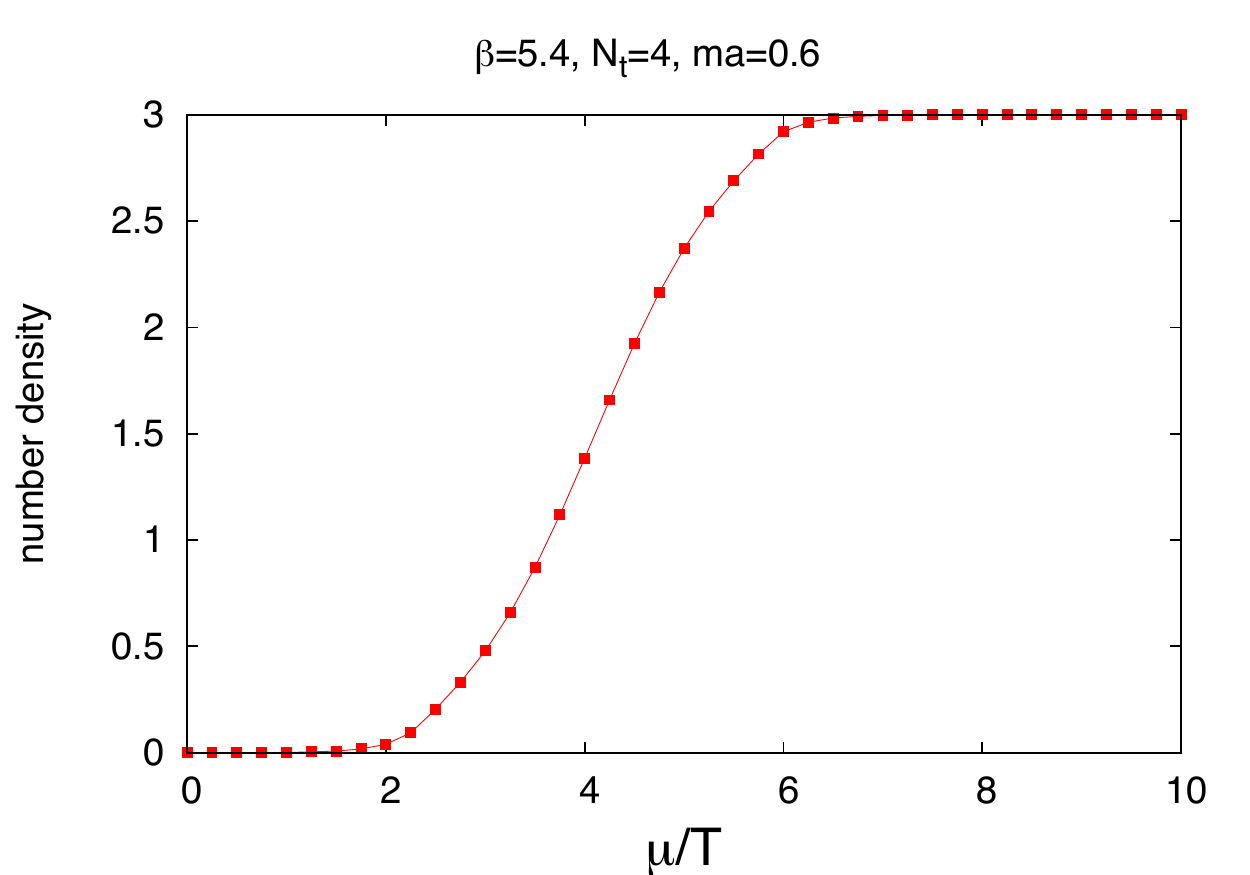}
}
\caption{Mean field solutions for Polyakov lines and number density at finite chemical potential at the Wilson
action and $N_t=4$ at $\b=5.04,~ ma=0.2$ (subfigures a,b) and $\b=5.4,~ma=0.6$ (subfigures b,c).}
\label{Wilsonmf}
\end{figure}
    
    In Fig. \ref{Wilsonmf} we display the mean-field results for the VEV of Polyakov lines $\tr U_\vx,~\tr U_\vx^\dg$, and the quark number density, vs.\ $\m/T$.  Note that the density has a plateau at number density $=3$, which is the expected limit for staggered fermions
(no rooting) on a lattice.  These results are qualitatively similar to what has been seen previously for heavy quarks, although the quark masses are well outside the range of validity of the hopping parameter expansion.  It is worth noting that the $h$ parameter in the effective action turns out to be quite small, equal to $h=0.033$ and $h=0.017$ in the two cases shown.

\begin{figure}[htb]
\centering
\subfigure[]  
{  
\label{LWuv1} 
 \includegraphics[scale=0.4]{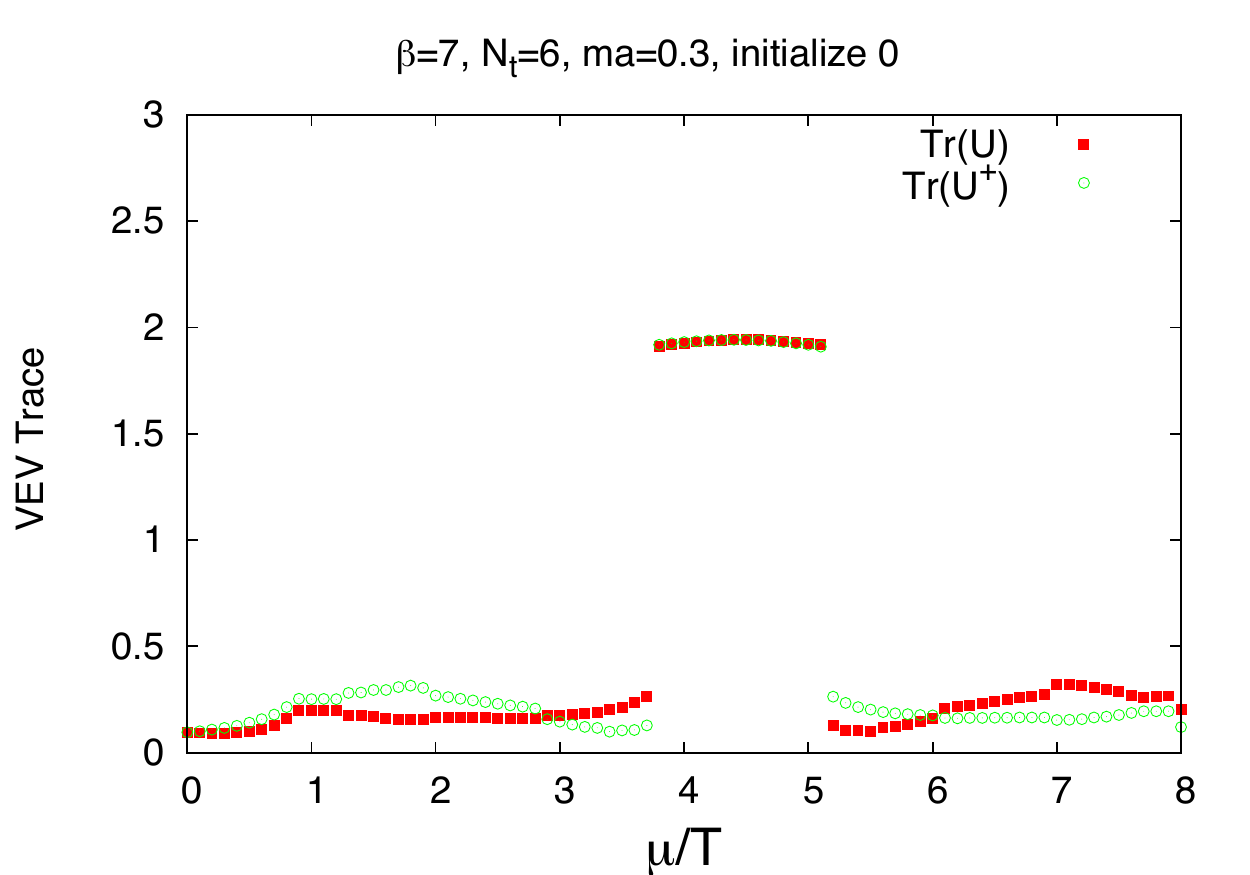}
}
\label{LWn1}
\subfigure[ ]  
{  
\label{LWuv2} 
 \includegraphics[scale=0.4]{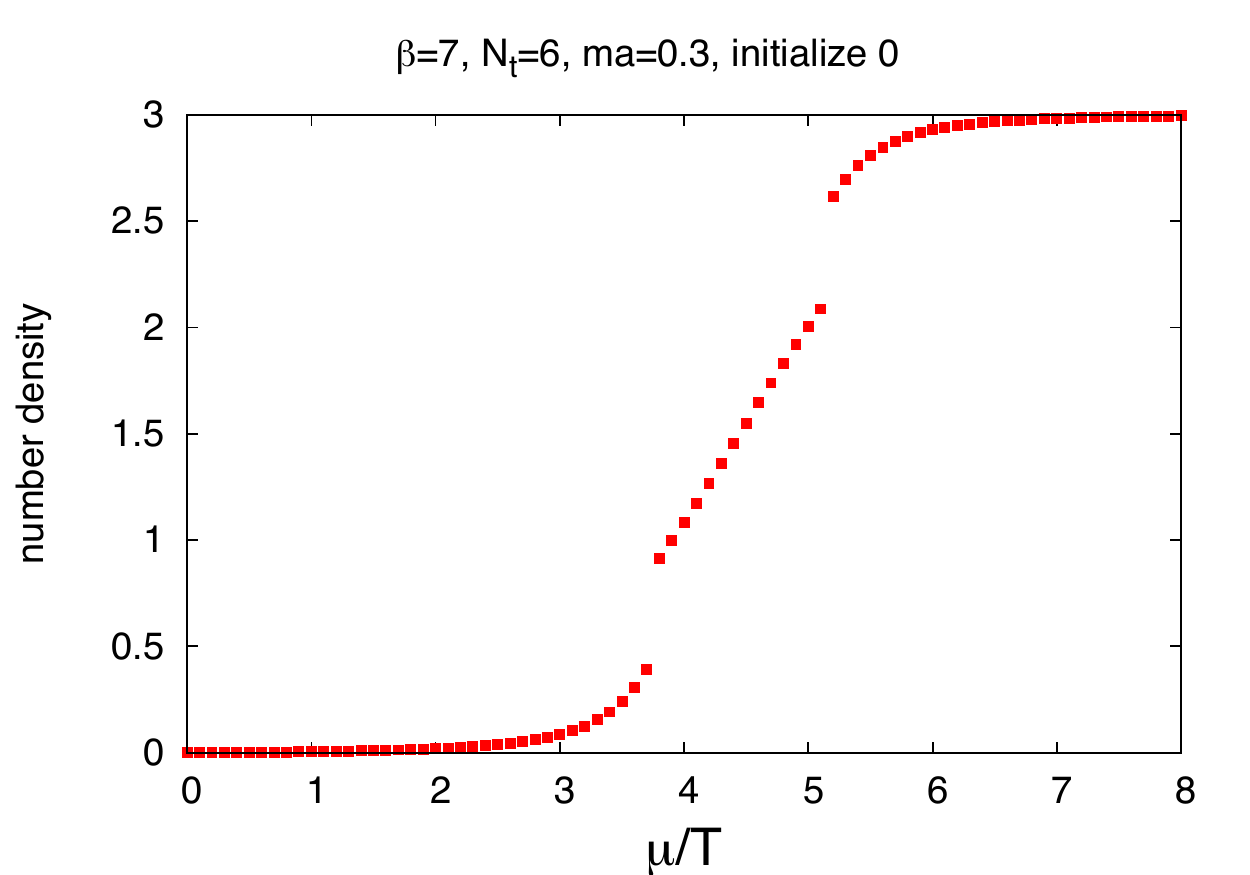}
}
\label{LWn2}
 \subfigure[]  
{   
 \includegraphics[scale=0.4]{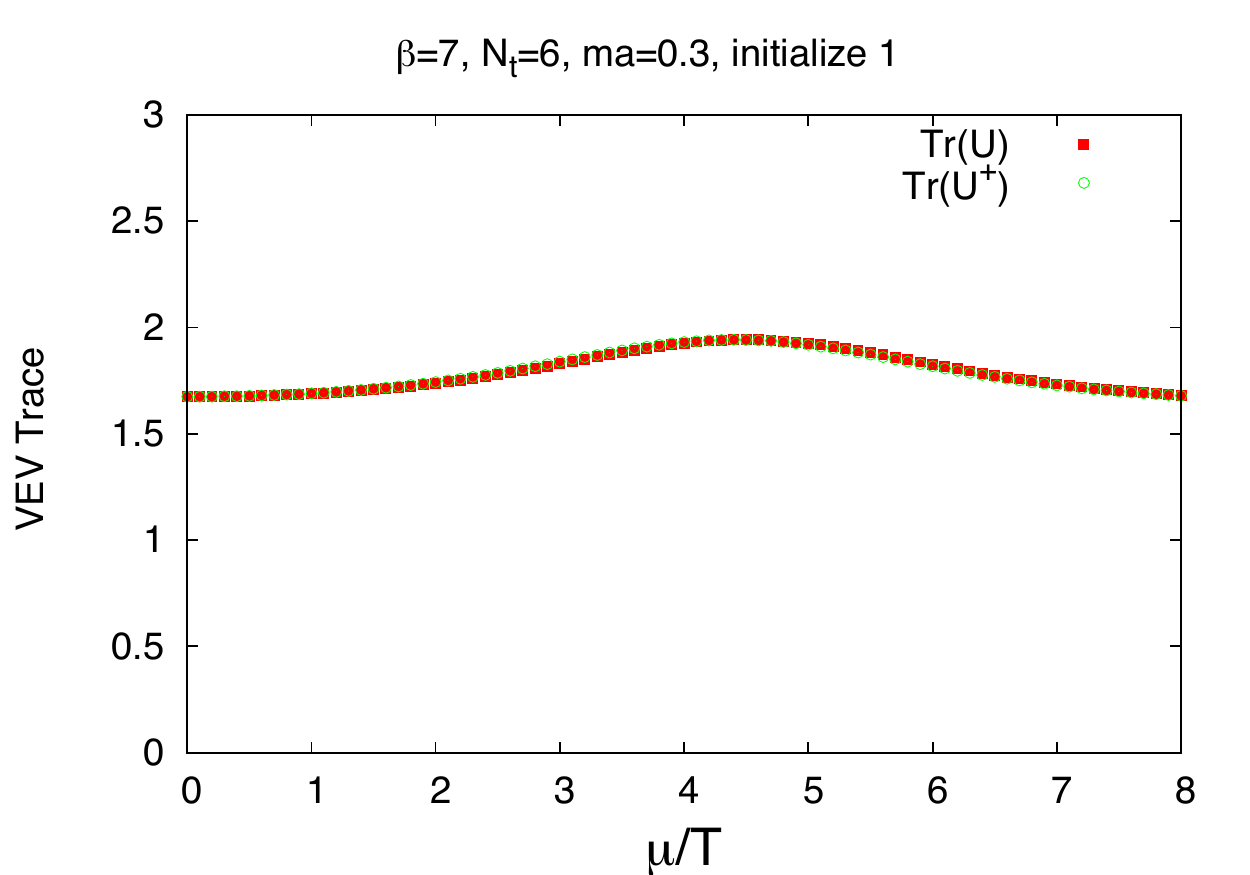}
}
\subfigure[ ]  
{   
 \includegraphics[scale=0.4]{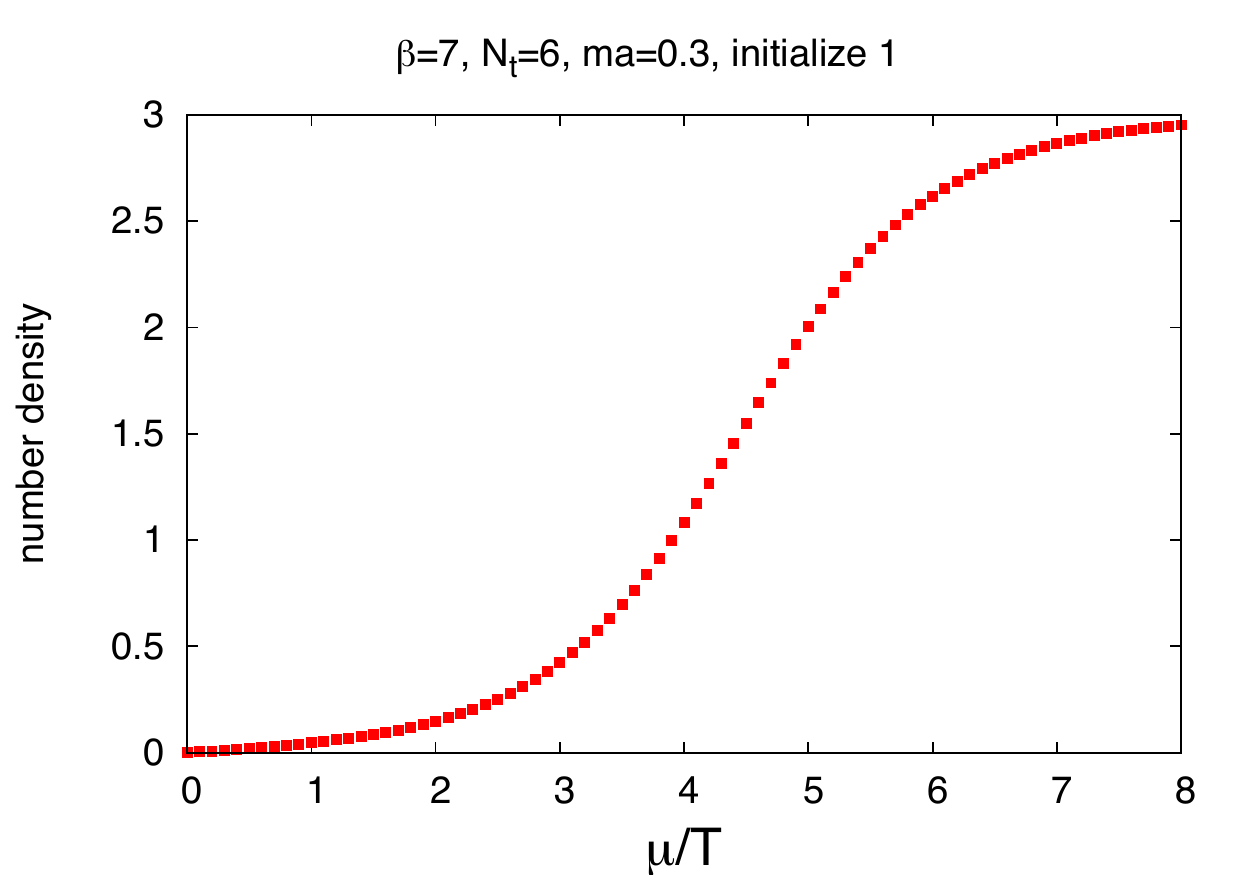}
}
\caption{Mean field solutions for Polyakov lines and number density vs.\ $\m$ in the PLA corresponding to a L\"uscher-Weisz action lattice gauge theory at $\b=7.0,~ma=0.3,~N_t=6$.  In subfigures (a,b) the routines look for a solution of the mean field equations closest to $u=v=0$, while in subfigures (c,d) the solution closest to $u=v=1$ is chosen.}
\label{LWmf}
\end{figure}

   The L\"uscher-Weisz case is more interesting.  Here there are different solutions of the mean-field consistency conditions, and if we
pick the solution with the smallest VEV for the Polyakov lines, then we find two phase transitions, seen in Figs.\ \ref{LWmf}(a) and \ref{LWmf}(b).  If instead we
pick the unphysical branch, the solution is as shown in Figs.\ \ref{LWmf}(c) and \ref{LWmf}(d).  Normally one would pick the solution with the lowest free energy, and this turns out to be the unphysical branch.  The solution with the slightly higher free energy corresponds, at $\m=0$, to the phase of the underlying lattice gauge theory, and we already know that this phase is stable for as long as we have carried out the Monte Carlo simulation.  So it cannot be discarded.

  The reliability of mean field theory can only be assessed where can compare our results with some other method, and so far this is only possible at $\m=0$, where we can compare Polyakov line values obtained from mean field theory and from lattice Monte Carlo simulations.
The comparison is shown in Table \ref{tab2}, and it is clear that there is excellent agreement.
\begin{table}[htb]
\begin{center}
\begin{tabular}{|c|c|c|c|c|c|} \hline
         action &  $ N_t $ & $\b$ & $ma$ &${1\over 3}\langle \tr U \rangle$& ${1\over 3}\langle \tr U \rangle_{mf}$ \\
\hline
        Wilson &          4 &  5.04 & 0.2  & 0.01778(3) & 0.01765   \\
        Wilson &          4 &  5.2  & 0.35 & 0.01612(4) & 0.01603    \\ 
        Wilson &          4 &  5.4  & 0.6   & 0.01709(5) &  0.01842  \\ 
L\"uscher-Weisz I&  6 &  7.0 & 0.3    & 0.03580(4) & 0.03212    \\   
L\"uscher-Weisz II& 6 &  7.0 & 0.3    & 0.554(1) & 0.5580    \\         
\hline
\end{tabular}
\caption{Polyakov line expectation values from numerical simulations of lattice gauge theory (column 5) , compared to mean field estimates (column 6).  For the L\"uscher-Weisz action there are multiple solutions of the mean field equations. The solution in L\"uscher-Weisz I is the one found by a search routine initialized at $u=v=0$, while the solution in L\"uscher-Weisz II corresponds to initialization at $u=v=1$. For L\"uscher-Weisz II, the value in column 5 was obtained from numerical simulation of the PLA, rather than the lattice gauge theory, with Polyakov lines initialized to 0.3.}
\label{tab2}
\end{center}
\end{table}
 
   To summarize, we have extended the relative weights method to dynamical staggered fermions in SU(3) lattice gauge theory.  The data is fit to a simple ansatz motivated by the heavy quark form, and at $\m=0$ we find good agreement between Polyakov line correlators computed in the effective action and the underlying lattice gauge theory.  The effective theory can be solved at $\m=0$ by a mean field technique.  It would be very interesting to compare our results with those of any other method that has been suggested for dealing with the sign problem.
   
    In one case we have encountered an extreme example of non-local couplings, and in this case the results of the simulation of the effective action depend on the starting point, i.e.\ there are very long-lived metastable states.  We will either need some criterion for selecting the correct phase in such cases, or else restrict the method to a region of parameter space where metastable states are not an issue.
    
    A more detailed presentation of the results outlined here can be found in \cite{Hollwieser:2016hne}.

\bibliography{pline} 
 

\end{document}